\documentclass[conference]{IEEEtran}
\IEEEoverridecommandlockouts
\usepackage{cite}
\usepackage{amsmath,amssymb,amsfonts}
\usepackage{algorithmic}
\usepackage{graphicx}
\usepackage{textcomp}
\usepackage{xcolor}
\def\BibTeX{{\rm B\kern-.05em{\sc i\kern-.025em b}\kern-.08em
    T\kern-.1667em\lower.7ex\hbox{E}\kern-.125emX}}
\begin{document}

\title{Quantum Software Engineering Challenges from Developers' Perspective: Mapping Research Challenges to the Proposed Workflow Model\\

\thanks{This work has been supported by the Academy of Finland (project DEQSE 349945) and Business Finland (Project TORQS 8582/31/2022, and Project FrameQ 8578/31/2022).}
}

\author{\IEEEauthorblockN{Majid Haghparast}
\IEEEauthorblockA{%\textit{dept. name of organization (of Aff.)} \\
\textit{University of Jyväskylä}\\
Jyväskylä, Finland \\
majid.m.haghparast@jyu.fi}
\and
\IEEEauthorblockN{Tommi Mikkonen}
\IEEEauthorblockA{%\textit{dept. name of organization (of Aff.)} \\
\textit{University of Jyväskylä}\\
Jyväskylä, Finland \\
tommi.j.mikkonen@jyu.fi}
\and
\IEEEauthorblockN{Jukka K. Nurminen}
\IEEEauthorblockA{%\textit{dept. name of organization (of Aff.)} \\
\textit{University of Helsinki}\\
Helsinki, Finland \\
jukka.k.nurminen@helsinki.fi}
\and
\IEEEauthorblockN{Vlad Stirbu}
\IEEEauthorblockA{%\textit{dept. name of organization (of Aff.)} \\
\textit{University of Jyväskylä}\\
Jyväskylä, Finland \\
vlad.a.stirbu@jyu.fi}

}

\maketitle

\begin{abstract}
Despite the increasing interest in quantum computing, the aspect of development to achieve cost-effective and reliable quantum software applications has been slow. One barrier is the software engineering of quantum programs, which can be approached from two directions. On the one hand, many software engineering practices, debugging in particular, are bound to classical computing. On the other hand, quantum programming is closely associated with the phenomena of quantum physics, and consequently, the way we express programs resembles the early days of programming. Moreover, much of the software engineering research today focuses on agile development, where computing cycles are cheap and new software can be rapidly deployed and tested, whereas in the quantum context, executions may consume lots of energy, and test runs may require lots of work to interpret. In this paper, we aim at bridging this gap by starting with the quantum computing workflow and by mapping existing software engineering research to this workflow. Based on the mapping, we then identify directions for software engineering research for quantum computing.
\end{abstract}


\section{Introduction}
Quantum computers (QC) outperform classical computers in different ways. Making Quantum computing accessible to the software developers calls for novel tools, methods, and processes that specifically concentrate on developing software systems on top of the quantum mechanics. There are a huge number of quantum software aspects that can be explored \cite{weder2020quantum,abreu2021quantum,gemeinhardt2021towards,serrano2022quantum}. Authors in \cite{weder2020quantum} and\cite{weder2022quantum} studied the quantum software lifecycle. In this paper, we study the quantum software engineering challenges from the developers' perspective and review a number of proposals that have so far been made for quantum software engineering. In summary, two different approaches can be identified. One approach is to build on top of agile and iterative software development (e.g. \cite{qp4se}) and experimentally solving data science problems (e.g. \cite{aho2020demystifying}) and related tooling (e.g. \cite{nguyen2022qfaas}). Another approach is to rely on formalisms, such as theorem proving and model checkers (e.g., \cite{liu2019formal}), to ensure that the composed quantum programs are constructed correctly. Instead of embracing either side, we propose viewing quantum software engineering as a new type of co-design problem, where iterative software development and formalisms complement each other to deliver reliable quantum applications. In addition, we frame software engineering research to test the proposed approach.

\section{Composing Quantum Software: An Engineering View}

\subsection{Quantum Software Workflow Simplified}

\begin{figure}
    \centering
    \includegraphics[width=1\columnwidth]{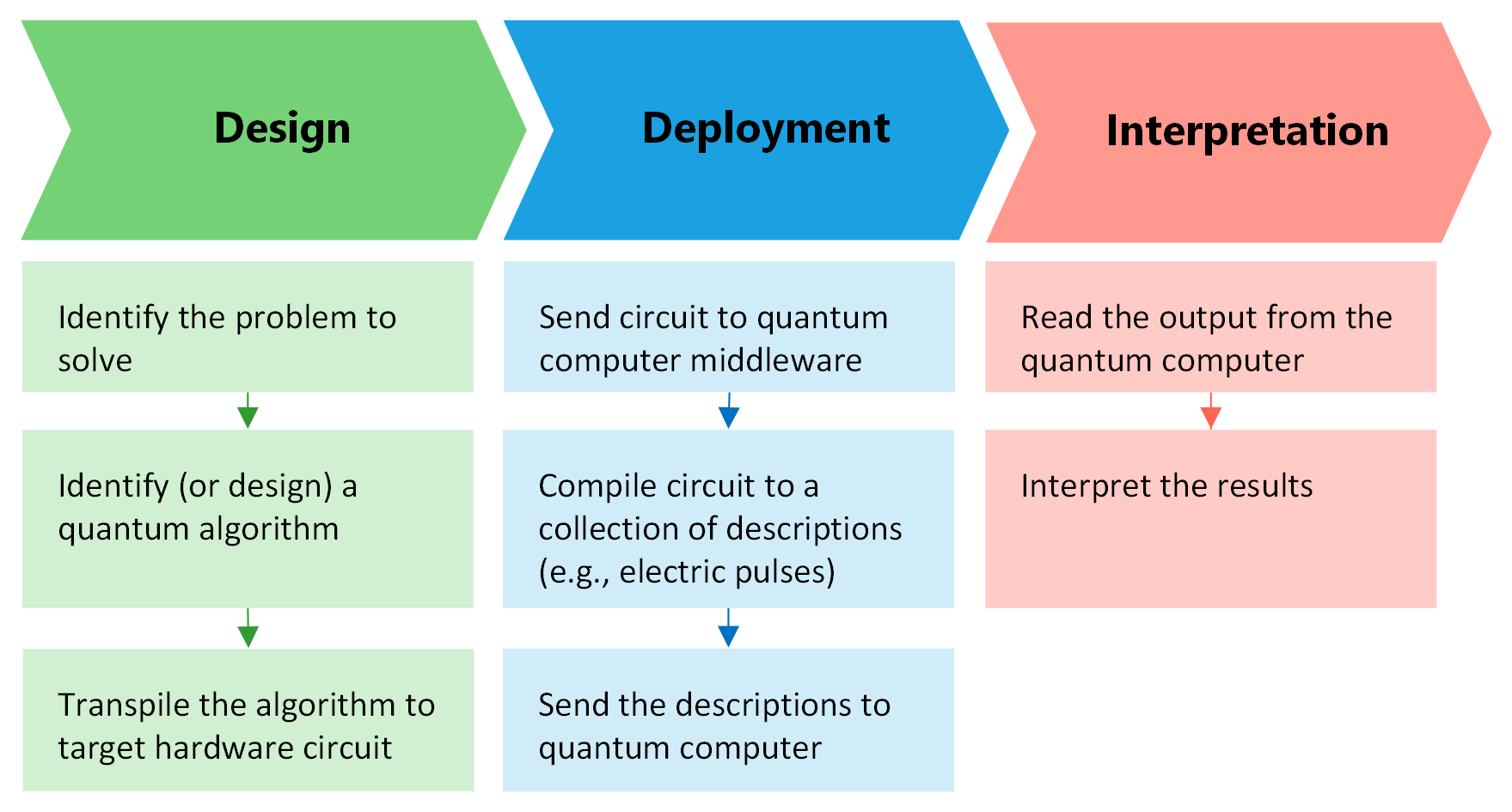}
    \caption{Steps required to design, deploy, and interpret quantum executions.}
    \label{fig:steps}
\end{figure}

In the following, we propose a quantum software engineering workflow based on the steps needed to run quantum programs. A quantum algorithm's journey from an idea to execution includes several steps, depicted in Figure \ref{fig:steps}, based on IQM Academy course launch\footnote{https://meetiqm.com/resources/press-releases/iqm-launches-free-online-course/, visited July 24, 2023.}. The steps are discussed in more detail below:
\subsubsection*{Composing Quantum Programs}  
\begin{itemize}
    \item Step 1: Identify a computational problem to solve.
    \item Step 2: Identify (or design) a quantum algorithm to solve the problem. Today, e.g., Python can be used to express the algorithm at various levels of abstraction. However, to run an algorithm, the level of abstraction corresponds to the quantum circuit consisting of gates (detailed in Step 3). While the available tools allow for designing new quantum algorithms, it is also possible to reuse already existing systems. Sometimes it is simpler to formulate the computational problem to a form that the quantum computer is able to serve. The number of known useful algorithms is small, and instead of crafting own designs, formulating the computational problem so that the quantum computer is able to serve it can be essential.
   \item Step 3: The circuit is Transpiled or converted to an equivalent version where only gates available in the corresponding quantum computer are used.
    \end{itemize}
\subsubsection*{Deploying and Running Quantum Programs}
\begin{itemize}
    \item Step 4: Circuit is sent to quantum computer middle-ware, possibly with hardware and calibration settings. 
    \item Step 5: The middle-ware then compiles the circuit to a collection of descriptions (e.g. electrical pulses in certain quantum hardware platforms)
    \item Step 6: The descriptions from step 5 are sent to the execution to the quantum chip.
\end{itemize}

\subsubsection*{Interpreting Quantum Execution Results}
\begin{itemize}
    \item Step 7: A measurement device reads the output produced (e.g. by the electrical pulses in certain quantum hardware platforms) and sends the results to the user as classical bits instead of measured values. Usually, the same program is run multiple times so that there will be statistical evidence for post-processing.
    \item Step 8: Post-processing software reads, analyses, and displays the results to the user, allowing one to consider the outcome of the execution and possibly return to Step 2 to debug the algorithm or the implementation. However, as the results of quantum executions are stochastic, debugging can be difficult because it may simply happen that at this execution time, something unlikely happened, and therefore the results are as they are. Moreover, quantum states cannot be studied while debugging because measurements change the quantum states. Hence, running the program step-by-step in a debugger to inspect the states is impossible. However, deducing quantum states from measured data under the term quantum state tomography \cite{cramer2010efficient} is possible at any time during the computation, but this procedure requires many runs and significant resources.

\end{itemize}

Clearly, there are lots of additional software needed in the process in addition to the actual quantum algorithm. Hence, many software engineering themes have emerged in the quantum computing domain, with ideas from software engineering borrowed in many forms. Below, we map ongoing research to the above workflow to identify research gaps.

\subsection{Design}
Quantum programming languages, are used to express quantum programs. The concrete design includes quantum algorithms and quantum circuits (low-level representation), which both describe quantum computations. Transpiling can also be a part of the refinements carried out as a part of the development. Iterative development, the mainstream development approach today, has also been proposed for quantum development, despite the fact that the full workflow proposed above can be too complex for frequent iterations. For instance, in \cite{qp4se}, the authors propose using iterations that rely on increasingly complex infrastructure, with concept development taking place with simulations at the developer's own computer, more complex simulations in the cloud, and only once the concept and the algorithm have been fixed, the real quantum computer is activated. Quantum programming languages, such as Qiskit\cite{Qiskit} and Cirq\cite{cirq_developers_2022_7465577}, have been proposed to simplify the development of quantum algorithms. Such languages typically introduce abstractions that reflect the properties of quantum circuits and hence simplify the development to some degree. In contrast, also extensions and libraries to existing systems have been proposed, building on top of, e.g., Python \cite{kaiser2021learn}. Existing languages mainly assume notations that reflect quantum circuits, whereas actual quantum algorithms are more abstract. Examples include Grover's algorithm \cite{jozsa1999searching}, realizing it on a given quantum computer needs special considerations, and human intelligence \cite{mandviwalla2018implementing}. By borrowing ideas from traditional computing, this step can be supported with model transformations \cite{sendall2003model} or methods such as templating techniques such as those proposed by Kellomäki and Mikkonen \cite{kellomaki2000design}. Model transformations and templating can work in two different ways. One is that a quantum algorithm is transformed into a known one for which an efficient implementation exists. Another is refinements that transform abstract algorithms towards quantum circuits. For the former, we trust that the body of knowledge in the field of classical algorithm design and the necessary machinery exists already. For the latter, we need new kinds of abstractions and refinement. Finally, formal methods for quantum computing \cite{liu2019formal}, quantum circuit design \cite{amy2019formal}, and model checking for communicating quantum processes \cite{davidson2012model} have been proposed . Moreover, classic systems such as Isabelle/HOL have been fine-tuned to work with quantum computations \cite{bordg2021certified}.

\subsection{Deployment}
Once completed, a quantum program is sent to execution to a quantum computer. A quantum algorithm is run (e.g., in terms of pulses fed to the quantum computer, for some specific QC platforms, with quantum circuits play defining how the pulses are generated). Today we have several techniques to deploy programs to execution. For instance, containers, simply sending instructions, or even quantum circuits are feasible solutions. The design of this deployment and execution middle-ware needs no special considerations associated with quantum computing, and traditional software design will satisfy the requirements. However, as the middle-ware directly interacts with the quantum computer, it cannot easily be shared across different quantum computers. In existing research, service-oriented computing (e.g., \cite{moguel2022quantum}) has been proposed to help deploy and run quantum programs. Indeed, there are many similarities in how service-oriented and actual quantum systems work -- for instance, IQM\footnote{https://www.meetiqm.com, visited July 24, 2023}, a quantum computer manufacturer, uses HTTP requests to send circuits and HTTP responses to return the results. Even more advanced systems have been proposed, such as serverless computing \cite{nguyen2022qfaas}. However, at this point, it is not immediately obvious what would be the actual benefit of added flexibility and configurability. In addition, more holistic research has been proposed for quantum computing. For instance, a recent systematic literature review \cite{khan2022software} studied architectural process, modeling notations, architecture design patterns, tool support, and challenging factors for quantum software architecture. Results of the SLR indicate that quantum software represents a new genre of software-intensive systems. However, existing processes and notations can be tailored to derive the architecting activities and develop modeling languages for quantum software. The results are echoed in \cite{ahmadtowards}, with a more process-oriented focus.

\subsection{Interpretation}
The results of quantum computations are stochastic in their nature. There are two sources of stochasticity. Firstly, quantum phenomena and resulting algorithms are genuinely unpredictable, but the probability of the right answer is high enough. Secondly, because of the unreliable operation of quantum computers, the quantum state can be corrupted, resulting in a random wrong answer. The interpretation between the two requires post-processing of the data, where the result is identified. Techniques for separating errors from the correct results seem specific to particular implementations, based on recent literature \cite{maciejewski2020mitigation,kiktenko2016post}. However, since it seems likely that the errors and other corrupted results are not only algorithm and domain-specific but also computer specific, there seems to be an opportunity to expand the infrastructure with computing device-specific error correction mechanisms.

\section{Research Agenda for Quantum Software}
\subsection{Design}
Today, a major problem in composing quantum software is the lack of understanding what quantum approach should be used to solve a problem at hand. So far, a set of quantum algorithms has been identified, but the problems where they can be applied still need to be understood. Hence, a taxonomy to consider whether and how a quantum computer outperforms classical ones when solving a problem would be a step forward. Another research direction for developing quantum software is enabling experimental, live prototyping \cite{swift2015live}, with facilities such as Read–Eval–Print loop \cite{ellis2019write}. For instance, while exploring a new algorithm, a coarse-grained simulation could be used, whereas, for production use, the full end-to-end quantum workflow would be run. This, in turn, could help better conceptualize quantum applications. In addition, tools (compilers) that map abstract quantum algorithms to low-level quantum circuits is an important direction for future research. This rehabilitation of formal methods, both in terms of proving the correctness of quantum software either manually and with theorem provers \cite{liu2019formal} and in terms of model checkers, seems to gather traction already. The rationale is that quantum computers are limited in capacity, and problems such as state explosion are not key concerns. However, since quantum logic differs from the classical one, it will need to adapt the tooling, which in itself can introduce new research directions. Finally, the correctness of the tooling itself needs attention. This point has been raised in \cite{burgholzer2020verifying}, where the authors propose a scheme for quantum circuit equivalence checking specialized to verify results of the IBM Qiskit quantum circuit compilation flow. Furthermore, using simulations to study the equivalence of quantum circuits has been proposed \cite{burgholzer2020power} to verify transformations that are a part of the process of refining quantum algorithms to quantum circuits fit for a particular quantum computer. Obviously, also deployment and interpreting the quantum results need similar correctness considerations.
\subsection{Deployment}
Deploying quantum software is largely business as usual. In the near term, programs are so small that no special infrastructure will be necessary, and existing cloud computing techniques will suffice to offer quantum computing as a service. Similarly, interfacing with the quantum computer requires low-level, real-time software. However, we can already compose such, and state-of-the-practice provides adequate tools for this. The part where we see room for research in quantum-related deployment is parametrization and scripting so that the developer can be liberated from some of the complications. While implementation techniques for such are readily available, we expect that these can introduce additional flexibility to quantum computing. Inspiration for such research can be seeked from serverless computing \cite{grossi2021serverless,nguyen2022qfaas} and service APIs \cite{garcia2021quantum} in quantum context, which are widely used in cloud systems.
Hybrid quantum-classical approaches require an efficient interplay between quantum and classical resources which needs attention for deployment\cite{leymann2021hybrid}. 

\subsection{Interpretation}
Clearly, interpreting quantum results is a key research topic. The first part of this problem is separating results and errors in stochastic outcomes of quantum executions. In analogy to ML fault tolerance patterns \cite{myllyaho2022misbehaviour}, we expect that the errors will form patterns. We propose studying these patterns from two directions. The first source of errors is related to the quantum algorithm and related circuits, which may be prone to certain types of errors. The other source of errors is the quantum computer at hand and the technology used to implement it, where executions may result in different error patterns. As the mechanism for detecting such errors, we propose data science methods in general and machine learning in particular, following the path shown by \cite{lloyd2013quantum} and \cite{yang2023survey}. Checking the applicability of quantum computation results is another direction for future research. For now, classical computers are powerful enough to check the results, but as the capacity of quantum computers increases, research will also be needed to verify and validate their outcomes with new techniques. Again, we expect that machine learning will enable advancement in this direction.
%machine learning in particular (see, e.g., \cite{lloyd2013quantum} and  \cite{yang2023survey}). 

\section{Conclusions}
Quantum computing has recently been gaining a lot of interest in the field of software engineering. Topics such as programming languages and, more generally, tools for composing quantum programs and deployment infrastructure to run quantum code have been addressed. However, to a degree, it can be argued that the research has originated from the viewpoint of established software engineering practices instead of the true needs of quantum computing practitioners. 
In this paper, we propose a research agenda based on the development, deployment, and use of quantum programs. Based on this, the practitioners' software engineering problems in the field of quantum computing today revolve around understanding the capabilities of quantum computing as well as interpreting the results. The former includes questions such as what quantum computing is good for, how to reuse existing quantum algorithms, and how to compose new ones reliably. The latter, in contrast, requires data engineering research to identify which results are reliable and which are erroneous in one way or another. 

\bibliographystyle{IEEEtran}
\bibliography{IEEEabrv,refs}

\end{document}